\documentclass[10pt,conference]{IEEEtran}

\usepackage{graphicx}
\usepackage{subcaption}
\usepackage{tikz}
\usepackage{pgfplots}
\usetikzlibrary{shapes.geometric}
\usetikzlibrary{positioning}
\usepackage[normalem]{ulem}
\usepackage{xspace}
\usepackage{listings}
\usepackage{booktabs}
\usepackage{tabularx}
\usepackage{caption}
\usepackage{subcaption}

\usepackage{hyperref}

\definecolor{codeblue}{rgb}{0.2,0.2,1}
\definecolor{codegreen}{rgb}{0,0.6,0}
\definecolor{codegray}{rgb}{0.5,0.5,0.5}
\definecolor{codepurple}{rgb}{0.58,0,0.82}
\definecolor{backcolour}{rgb}{0.95,0.95,0.92}
\lstdefinestyle{mystyle}{
    commentstyle=\color{codegreen},
    keywordstyle=\color{codeblue}\bfseries,
    numberstyle=\tiny\color{codegray},
    stringstyle=\color{codepurple},
    basicstyle=\footnotesize\ttfamily,
    breakatwhitespace=false,         
    breaklines=true,                 
    captionpos=b,                    
    keepspaces=true,                 
    numbers=none,                    
    numbersep=0pt,                  
    showspaces=false,                
    showstringspaces=false,
    showtabs=false,
    frame=single,
    tabsize=4,
    columns=flexible
}

\begin{document}

\title{Automated Design Space Exploration of CGRA Processing Element Architectures using Frequent Subgraph Analysis}
\author{
\IEEEauthorblockN{Jackson Melchert\IEEEauthorrefmark{1},
Kathleen Feng\IEEEauthorrefmark{1},
Caleb Donovick\IEEEauthorrefmark{2},
Ross Daly\IEEEauthorrefmark{2},
Clark Barrett\IEEEauthorrefmark{2},
Mark Horowitz\IEEEauthorrefmark{2}\IEEEauthorrefmark{1}, \\
Pat Hanrahan\IEEEauthorrefmark{2}\IEEEauthorrefmark{1}, and
Priyanka Raina\IEEEauthorrefmark{1}}
\IEEEauthorblockA{\IEEEauthorrefmark{1}Dept. of Electrical Engineering, \IEEEauthorrefmark{2}Dept. of Computer Science\\
Stanford University, Stanford, CA}}

\maketitle

\thispagestyle{empty}

\begin{abstract}

The architecture of a coarse-grained reconfigurable array (CGRA) processing element (PE) has a significant effect on the performance and energy efficiency of an application running on the CGRA. This paper presents an automated approach for generating specialized PE architectures for an application or an application domain. Frequent subgraphs mined from a set of applications are merged to form a PE architecture specialized to that application domain. For the image processing and machine learning domains, we generate specialized PEs that are up to 10.5$\times$ more energy efficient and consume 9.1$\times$ less area than a baseline PE.

\end{abstract}

\section{Introduction}

Coarse-grained reconfigurable arrays (CGRAs) have been widely studied in recent years and serve as a midpoint between the flexibility of an FPGA and the performance and energy efficiency of an ASIC~\cite{artem}. A CGRA can achieve better performance and energy efficiency than an FPGA because its processing elements (PEs) have specialized arithmetic units that operate at a word-level, rather than a bit-level, granularity. They have better reconfigurability than an ASIC due to a flexible interconnect and configurable PEs. However, CGRAs do not constitute just one point in the spectrum between FPGAs and ASICs; rather, they occupy a large design space from specialized and efficient CGRAs to flexible CGRAs that can accelerate many applications. CGRAs are useful when designed with an application domain in mind, allowing for appropriately specialized PE and memory architectures while leaving some flexibility to accommodate the evolution of the applications in the domain.

The design of CGRA PEs has a direct and significant effect on application mapping and CGRA power, performance, and area while also having an indirect effect on the CGRA interconnect. In traditional island-style FPGAs, a significant portion of the die area is dedicated to interconnect and configuration~\cite{Farooq2012}. CGRAs reduce this overhead by increasing compute density~\cite{artem}. To achieve even higher compute density, design space exploration (DSE) of specialized CGRA PE architectures is required.

In this paper, we present a methodology for the design space exploration of CGRA PEs. Our methodology is based on analyzing applications within an application domain to find common computational blocks that can be easily accelerated with specialized PEs. Using this methodology, we explore the space between general CGRAs that can execute many applications and specialized CGRAs that execute a more limited set but achieve high performance and energy efficiency.

Our framework encompasses application analysis, PE specification, CGRA hardware generation, and compiler creation in an easy-to-use toolchain that requires very little manual effort. Given an application or a set of applications, the framework can produce many PE architectures that explore the design space along with the rest of the CGRA and the compiler to evaluate the PE designs. Our contributions are as follows:

\begin{enumerate}
    \item Develop a methodology to analyze applications using subgraph mining and generate candidate PEs specialized for those applications using subgraph merging.
    \item Automatically generate CGRAs with specialized PEs along with a compiler to run applications on those CGRAs utilizing an agile hardware flow.
    \item Evaluate the area, energy, and performance of a CGRA running a variety of applications using PEs specialized to those applications or application domains.
\end{enumerate}

The rest of this paper is organized as follows: Section~\ref{design-space} introduces the design space axes of CGRA PEs. Section~\ref{application-analysis} describes our application analysis and PE specialization methodology. Section~\ref{exploration-framework} presents our DSE framework. Finally, Section~\ref{results} demonstrates the efficiency improvements obtained by specializing PEs using our framework for image processing and machine learning (ML) domains.

\section{Processing Element Design Space}
\label{design-space}

A survey of existing CGRA PE architectures indicates the presence of three design spaces axes~\cite{reconfig-computing-archs} --- the number and type of operations within the PE, the intraconnect of the PE, and the number of inputs to and outputs from the PE (Fig.~\ref{fig:dse-morphosys}a). 

\subsection{Number and Type of Operations}

\begin{figure}
    \centering
    \includegraphics[width=0.5\textwidth]{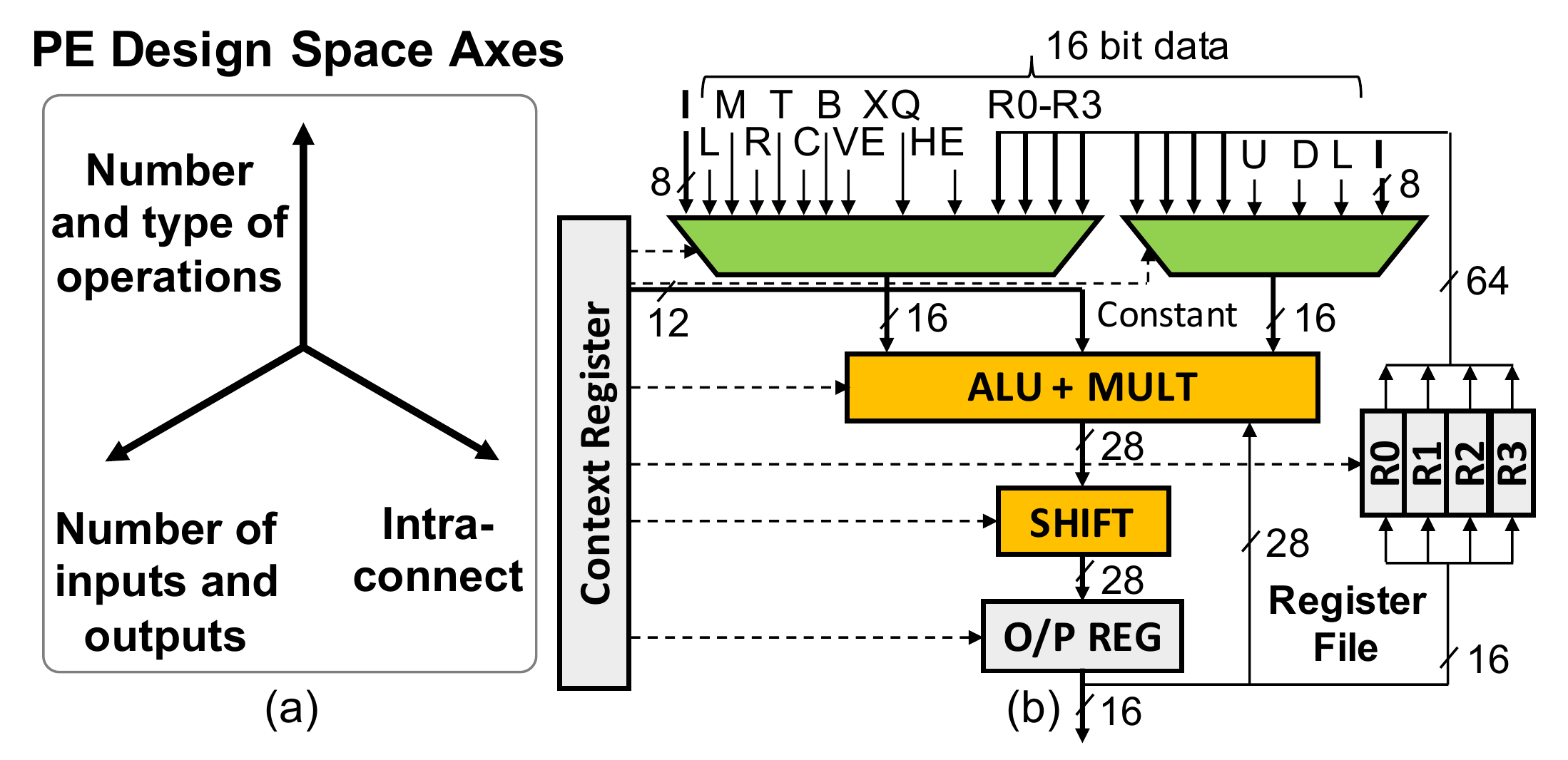}
    \caption{(a) Processing element (PE) design space axes. (b) A PE from MorphoSys CGRA (details omitted)~\cite{morphosys}.}
    \label{fig:dse-morphosys}
    \vspace{-0.5cm}
\end{figure}

CGRA PE architectures contain arithmetic units that can execute a number of operations, enabling a wide variety of applications to run on the CGRA. Typically, each PE contains at least one arithmetic unit that implements a handful of primitive operations such as add, subtract, shift, and comparisons. In addition, PEs may contain a multiplier and logic unit that can do bit operations. 
At one end of this design space axis are the most general PEs that contain one ALU and one multiplier. An example is the MorphoSys CGRA PE~\cite{morphosys} shown in Fig.~\ref{fig:dse-morphosys}b. This type of design lends itself to high utilization and similar efficiency no matter the application running on the CGRA. On the other end of this design space axis are more specialized PEs that have multiple ALUs or multipliers. For example, CGRAs targeting image processing or ML are likely to have the ability to do a vector multiply-add operation~\cite{simba}. Some complex PEs~\cite{plasticine} contain floating point logic that, while expensive, can enable running a wider range of applications. 

\subsection{Intraconnect}

The number of configurable paths through a PE affects its area and power, as well as  generality. A PE that can be configured to do a larger number of operations and more complex operations will have higher area and power, but it may require far fewer PEs to map to an application. On one end of this axis are PEs that have very few configurable paths. Since CGRA PEs can execute more than one operation, at least one multiplexer is needed to route the desired operation to the PE output. On the other end of this axis are PEs that have many multiplexers that enable many different configurable paths through the PE. A common example of this is to have multiplexers at the inputs of each arithmetic unit. Each input to the PE can be routed to either input of the ALU, so non-commutative operations like shifts can be achieved regardless of the order of the operands into the PE.

\subsection{Number of Inputs and Outputs}

\begin{figure}
    \centering
    \includegraphics[width=0.5\textwidth]{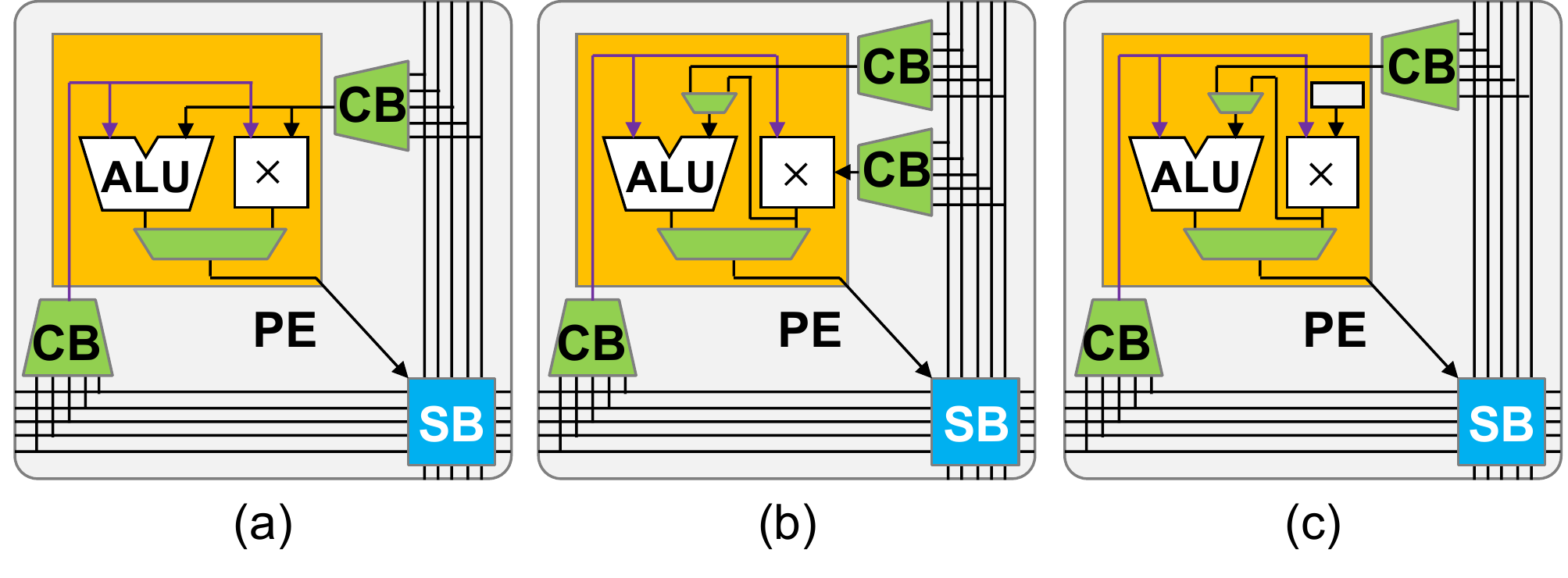}
    \caption{PE I/O: (a) PE with 2 inputs (2 CBs) and 1 output, (b) PE with 3 inputs (3 CBs), (c) Restricting the PE from (b) to receive one input from a constant register to reduce I/O.}
    \label{fig:cgra_pe_tile}
    \vspace{-0.5cm}
\end{figure}

The number of inputs and outputs (I/O) to each PE directly affects the size and number of connection boxes (CBs) and switch boxes (SBs) in the CGRA. The CBs take inputs from the routing tracks and feed them into PEs, while the SBs take outputs from the PE and route them to other PEs. As these components of the interconnect have high area and power costs, minimizing the I/O to the PE is critical for achieving an efficient CGRA. On one end of this design space axis are PEs that have two inputs and one output (Fig.~\ref{fig:cgra_pe_tile}a). This enables most arithmetic operations and results in small per-PE interconnect overhead. On the other end are designs that have many inputs to enable more complex operations, for example, a three input operation like a fused multiply add (Fig.~\ref{fig:cgra_pe_tile}b). 

Another aspect of this design space axis is reducing the number of inputs to the PE using constant registers. In many image processing applications, convolutions are done using a kernel with fixed weights. A PE that has a fused multiply-add to implement this convolution would usually need three inputs, however, one of those inputs, the kernel weight, could be replaced by a constant register (Fig.~\ref{fig:cgra_pe_tile}c). This register can then be given a value during the configuration of the CGRA, reducing the overhead of the interconnect. 

\section{Application Analysis}
\label{application-analysis}

All three design space axes described in Section~\ref{design-space} have many potential values, thus a na\"{i}ve exploration of the design space would lead to many candidate PEs. An intelligent method for exploring this design space and selecting interesting candidate PEs is needed for efficient exploration. In this section, we introduce our application analysis methodology, which is based on frequent subgraph mining and analysis.

\subsection{Subgraph Mining}

We start with an application written in Halide~\cite{halide2013_journal}, a domain-specific language (DSL) for image processing and machine learning. We use the Halide compiler from~\cite{dac} to lower the application to a CoreIR~\cite{coreir} dataflow graph containing compute and memory nodes. Since the compute nodes are primitive operations such as add, multiply, etc. and the edges are connections between those operations, frequent subgraphs of the application represent common (potentially complex) operations in the application. We use the frequent subgraphs as a starting point for constructing interesting PEs.

Finding, or mining, frequent subgraphs relies on the computation of subgraph isomorphisms which is an NP-complete problem. As frequent subgraph mining is very useful in a wide variety of fields, this is a well-researched problem. We use GRAMI~\cite{grami}, a subgraph mining tool for single large graphs. 

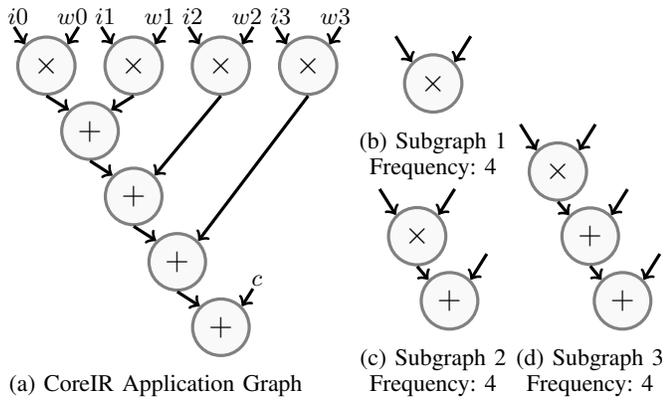
\begin{figure}
\centering
\captionsetup{justification=centering}
\tabskip=0pt
\valign{#\cr
  \hbox{%
    \begin{subfigure}[b]{.23\textwidth}
    \centering
        \begin{tikzpicture}[scale=0.29, font=\small,roundnode/.style={circle, draw=gray, fill=gray!5, very thick, minimum size=4mm},inputnode/.style={circle, draw=none,very thick, minimum size=4mm}]

            \node[] at (-0.3,6.25) {$i0$};
            \node[] at (3.7,6.25) {$i1$};
            \node[] at (7.7,6.25) {$i2$};
            \node[] at (11.75,6.25) {$i3$};
        
            \node[] at (2.2,6.25) {$w0$};
            \node[] at (6.2,6.25) {$w1$};
            \node[] at (10.2,6.25) {$w2$};
            \node[] at (14.25,6.25) {$w3$};
            
            \node[] at (10.65,-5.85) {$c$};
        
            \node[roundnode] at (1,4) (mul0) {\large$\times$};
            \node[roundnode] at (5,4) (mul1) {\large$\times$};
            \node[roundnode] at (9,4) (mul2) {\large$\times$};
            \node[roundnode] at (13,4) (mul3) {\large$\times$};
            
            \node[roundnode] at (3, 1) (add0) {\large$+$};
            \node[roundnode] at (5, -2) (add1) {\large$+$};
            \node[roundnode] at (7, -5) (add2) {\large$+$};
            \node[roundnode] at (9, -8) (add3) {\large$+$};

            \draw[very thick, <-] (mul0.north west) -- ++ (-.5, +.8);
            \draw[very thick, <-] (mul0.north east) -- ++ (+.5, +.8);
        
            \draw[very thick, <-] (mul1.north west) -- ++ (-.5, +.8);
            \draw[very thick, <-] (mul1.north east) -- ++ (+.5, +.8);
                    
            \draw[very thick, <-] (mul2.north west) -- ++ (-.5, +.8);
            \draw[very thick, <-] (mul2.north east) -- ++ (+.5, +.8);
            
            \draw[very thick, <-] (mul3.north west) -- ++ (-.5, +.8);
            \draw[very thick, <-] (mul3.north east) -- ++ (+.5, +.8);
            
            \draw[very thick, ->] (mul0.south) -- (add0.north west);
            \draw[very thick, ->] (mul1.south) -- (add0.north east);
            \draw[very thick, ->] (add0.south) -- (add1.north west);
            \draw[very thick, ->] (mul2.south) -- (add1.north east);
            \draw[very thick, ->] (add1.south) -- (add2.north west);
            \draw[very thick, ->] (add2.south) -- (add3.north west);
            \draw[very thick, ->] (mul3.south) -- (add2.north east);
            \draw[very thick, <-] (add3.north east) -- ++ (+.5, +.8);
        
    \end{tikzpicture}
    \caption{CoreIR Application Graph}
    \label{fig:subgraph_mining_app}
    \end{subfigure}%
  }\cr
  \noalign{\hfill}
  \hbox{%
    \begin{subfigure}{.12\textwidth}
    \centering
    \begin{tikzpicture}[scale=0.43,font=\small,roundnode/.style={circle, draw=gray, fill=gray!5, very thick, minimum size=5mm},inputnode/.style={circle, draw=none,very thick, minimum size=5mm}]
    
        \node[inputnode] at (0,6) (const0) {};
        \node[inputnode] at (2,6) (in0) {};
        
        \node[roundnode] at (1,4) (mul0) {\large$\times$};

        \draw[very thick, <-] (mul0.north west) -- ++ (-.5, +.8);
        \draw[very thick, <-] (mul0.north east) -- ++ (+.5, +.8);

    \end{tikzpicture}
    \caption{Subgraph 1 Frequency: 4}
    \label{fig:subgraph_mining_subgraph_1}
    \end{subfigure}%
  }\vfill
  \hbox{%
    \begin{subfigure}{.12\textwidth}
    \centering
    \begin{tikzpicture}[scale=0.43,font=\small,roundnode/.style={circle, draw=gray, fill=gray!5, very thick, minimum size=5mm},inputnode/.style={circle, draw=none,very thick, minimum size=5mm}]
    
        \node[roundnode] at (1,4) (add0) {\large$\times$};

        \node[roundnode] at (2, 2) (add1) {\large$+$};
        
        \draw[very thick, <-] (add0.north west) -- ++ (-.5, +.8);
        \draw[very thick, <-] (add0.north east) -- ++ (+.5, +.8);

        \draw[very thick, ->] (add0.south) -- (add1.north west);
        \draw[very thick, <-] (add1.north east) -- ++ (+.5, +.8);

    \end{tikzpicture}
    \caption{Subgraph 2 Frequency: 4}
    \label{fig:subgraph_mining_subgraph_2}
    \end{subfigure}%
  }\cr
  \vfill
  \hbox{%
    \begin{subfigure}{.11\textwidth}
    \centering
        \begin{tikzpicture}[scale=0.43, font=\small,roundnode/.style={circle, draw=gray, fill=gray!5, very thick, minimum size=5mm},inputnode/.style={circle, draw=none,very thick, minimum size=5mm}]
    
        \node[roundnode] at (0,6) (mul0) {\large$\times$};

        \node[roundnode] at (1,4) (add0) {\large$+$};

        \node[roundnode] at (2, 2) (add1) {\large$+$};
        
        \draw[very thick, <-] (mul0.north west) -- ++ (-.5, +.8);
        \draw[very thick, <-] (mul0.north east) -- ++ (+.5, +.8);
        
        \draw[very thick, ->] (mul0.south) -- (add0.north west);
        \draw[very thick, <-] (add0.north east) -- ++ (+.5, +.8);

        \draw[very thick, ->] (add0.south) -- (add1.north west);
        \draw[very thick, <-] (add1.north east) -- ++ (+.5, +.8);

    \end{tikzpicture}
    \caption{Subgraph 3 Frequency: 4}
    \label{fig:subgraph_mining_subgraph_3}
    \end{subfigure}%
  }\cr
}

\caption{Frequent subgraph mining on a convolution.}\label{fig:subgraph_mining}
\vspace{-0.5cm}
\end{figure}

GRAMI takes the application graph as an input in addition to a minimum number of times a subgraph can appear to be considered frequent. Fig.~\ref{fig:subgraph_mining} gives an example of frequent subgraph mining for a simple application. Fig.~\ref{fig:subgraph_mining_app} shows the CoreIR graph representation of a convolution $(((((i0 \times w0) + (i1 \times w1)) + (i2 \times w2)) + (i3 \times w3)) + c)$. Some frequent subgraphs of this application are in Fig.~\ref{fig:subgraph_mining_subgraph_1}, ~\ref{fig:subgraph_mining_subgraph_2}, and ~\ref{fig:subgraph_mining_subgraph_3}.

A subgraph of a CoreIR application can be interpreted as a PE architecture. Each operation in the subgraph has a hardware interpretation, so constructing PEs from many different subgraphs can be easily automated. However, not all frequent subgraphs are as interesting as they initially seem. Fig.~\ref{fig:subgraph_mining_subgraph_3} is an example of such a case. While the frequency is four, the occurrences overlap, and therefore, only two of the occurrences can be effectively accelerated. We use maximal independent set analysis to mitigate this issue.

\subsection{Maximal Independent Set Analysis}

Subgraphs that overlap in the application graph cannot be accelerated with fully utilized PEs that have the architecture to accelerate that subgraph. Fig.~\ref{fig:subgraph_mining_subgraph_3} shows a frequent subgraph of a convolution application. This subgraph has many occurrences in the application graph, although several of these occurrences overlap. If a PE had the architecture of this subgraph and was used to accelerate this application, it would result in underutilized PEs. 

One method to find when this problem occurs is maximal independent set analysis~\cite{maxindset}, which uses the following steps:

\begin{enumerate}
    \item Represent each occurrence of the subgraph in the application as a node in a new graph.
    \item Represent overlapping subgraphs as edges between nodes. Overlapping subgraphs are those whose occurrences share any node.
    \item Calculate the maximal independent set of this new graph; the size of this set is the number of times the subgraph exists in the application without overlaps.
\end{enumerate}

An independent set of a graph is a set of vertices in that graph which do not share a neighbor. A maximal independent set (MIS) is an independent set which cannot be grown by adding more vertices to it.
The size of MIS represents the number of fully utilized PEs with the architecture derived from the subgraph that can be used to run the application. 

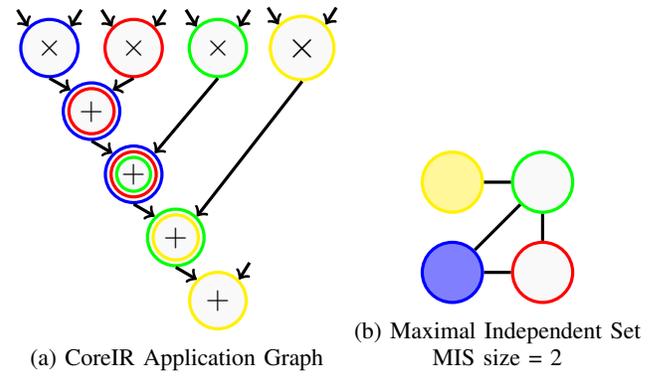
\begin{figure}
    \begin{centering}
    \captionsetup{justification=centering}
    \begin{subfigure}[b]{.25\textwidth}
        \centering
        \begin{tikzpicture}[scale=0.28, font=\small,roundnode/.style={circle, draw=gray, fill=gray!5, very thick, minimum size=7mm},doublenode/.style={circle, draw=none,very thick, minimum size=6mm},triplenode/.style={circle, draw=none,very thick, minimum size=4.5mm}]

            \node[roundnode, draw=blue] at (1,4) (mul0) {\large$\times$};
            \node[roundnode, draw=red] at (5,4) (mul1) {\large$\times$};
            \node[roundnode, draw=green] at (9,4) (mul2) {\large$\times$};
            \node[roundnode, draw=yellow] at (13,4) (mul3) {\Large$\times$};
            
            \node[roundnode, draw=blue] at (3, 1) (add0) {\large$+$};
            \node[doublenode, draw=red] at (3, 1) () {};
            \node[roundnode, draw=blue] at (5, -2) (add1) {\large$+$};
            \node[doublenode, draw=red] at (5, -2) () {};
            \node[triplenode, draw=green] at (5, -2) () {};
            \node[roundnode, draw=green] at (7, -5) (add2) {\large$+$};
            \node[doublenode, draw=yellow] at (7, -5) () {};
            \node[roundnode, draw=yellow] at (9, -8) (add3) {\large$+$};

            \draw[very thick, <-] (mul0.north west) -- ++ (-.5, +.8);
            \draw[very thick, <-] (mul0.north east) -- ++ (+.5, +.8);
        
            \draw[very thick, <-] (mul1.north west) -- ++ (-.5, +.8);
            \draw[very thick, <-] (mul1.north east) -- ++ (+.5, +.8);
                    
            \draw[very thick, <-] (mul2.north west) -- ++ (-.5, +.8);
            \draw[very thick, <-] (mul2.north east) -- ++ (+.5, +.8);
            
            \draw[very thick, <-] (mul3.north west) -- ++ (-.5, +.8);
            \draw[very thick, <-] (mul3.north east) -- ++ (+.5, +.8);
            
            \draw[very thick, ->] (mul0.south) -- (add0.north west);
            \draw[very thick, ->] (mul1.south) -- (add0.north east);
            \draw[very thick, ->] (add0.south) -- (add1.north west);
            \draw[very thick, ->] (mul2.south) -- (add1.north east);
            \draw[very thick, ->] (add1.south) -- (add2.north west);
            \draw[very thick, ->] (add2.south) -- (add3.north west);
            \draw[very thick, ->] (mul3.south) -- (add2.north east);
            \draw[very thick, <-] (add3.north east) -- ++ (+.5, +.8);
            
        \end{tikzpicture}
        \caption{CoreIR Application Graph}
        \label{fig:max_ind_set_app}

    \end{subfigure}%
    \begin{subfigure}[b]{.22\textwidth}
        \centering
        \begin{tikzpicture}[scale=0.3, font=\small,roundnode/.style={circle, draw=gray, fill=gray!5, very thick, minimum size=8mm}]

            \node[roundnode, draw=blue, fill=blue!50] at (0,0) (blue) {};
            \node[roundnode, draw=red] at (4,0) (red) {};
            \node[roundnode, draw=green] at (4,4) (green) {};
            \node[roundnode, draw=yellow, fill=yellow!50] at (0,4) (yellow) {};
    
            \draw[very thick, -] (blue) -- (red);
            \draw[very thick, -] (red) -- (green);
            \draw[very thick, -] (green) -- (blue);
            \draw[very thick, -] (green) -- (yellow);
        \end{tikzpicture}
        \caption{Maximal Independent Set MIS size = 2}
        \label{fig:max_ind_set_graph}

    \end{subfigure}%
    \caption{Maximal independent set of a frequent subgraph.}\label{fig:max_ind_set}
    \vspace{-0.5cm}
    \end{centering}
\end{figure}

Fig.~\ref{fig:max_ind_set} illustrates this on a simple application graph using the subgraph from Fig.~\ref{fig:subgraph_mining_subgraph_3}. The first occurrence of the subgraphs is outlined in blue in Fig.~\ref{fig:max_ind_set_app}, the second in red, the third in green, and the fourth in yellow. Each of these occurrences has a corresponding node in Fig.~\ref{fig:max_ind_set_graph}. The maximal independent set is the set of nodes that are filled with their respective color (yellow and blue). In this example, MIS size is 2, meaning this subgraph occurs twice in the application graph not including overlapping occurrences. The MIS size of a subgraph indicates how many fully utilized PEs that implement this subgraph can be used to accelerate the application. Using the size of this set, we have a better idea of which application subgraphs might be interesting starting points for PE architectures. 

\subsection{Subgraph Merging}

\begin{figure}
\begin{centering}

    \begin{subfigure}[b]{.15\textwidth}
        \centering
        \begin{tikzpicture}[scale=0.57, font=\small,roundnode/.style={circle, draw=gray, fill=gray!5, very thick, minimum size=9mm,inner sep=1pt}]
    
            \node[roundnode, align=center] at (0,2) (const) {const\\a0};
            \node[roundnode, align=center] at (2,2) (alu0) {\large$+$\\a2};
            \node[roundnode, align=center] at (1,0) (alu1) {\large$+$\\a1};
    
            \draw[very thick, <-] (alu0.north west) -- ++ (-.3, +.6);
            \draw[very thick, <-] (alu0.north east) -- ++ (+.3, +.6);
            \draw[very thick, ->] (alu0.south) -- (alu1.north east);
            \draw[very thick, ->] (const.south) -- (alu1.north west);
    
        \end{tikzpicture}
        \caption{Subgraph 1}
        \label{fig:subgraph_1}

    \end{subfigure}%
    \begin{subfigure}[b]{.15\textwidth}
        \centering
        \begin{tikzpicture}[scale=0.57, font=\small,roundnode/.style={circle, draw=gray, fill=gray!5, very thick, minimum size=9mm,inner sep=1pt}]
    
            \node[roundnode, align=center] at (-1,4) (const) {const\\b0};
            \node[roundnode, align=center] at (0,2) (mul) {\large$\times$\\b1};
            \node[roundnode, align=center] at (2,2) (alu0) {\large$+$\\b3};
            \node[roundnode, align=center] at (1,0) (alu1) {\large$+$\\b2};
    
            \draw[very thick, <-] (alu0.north west) -- ++ (-.3, +.6);
            \draw[very thick, <-] (alu0.north east) -- ++ (+.3, +.6);
            \draw[very thick, <-] (mul.north east) -- ++ (+.3, +.6);
            \draw[very thick, ->] (mul.south) -- (alu1.north west);
            \draw[very thick, ->] (alu0.south) -- (alu1.north east);
            \draw[very thick, ->] (const.south) -- (mul.north west);
    
        \end{tikzpicture}
        \caption{Subgraph 2}
        \label{fig:subgraph_2}

    \end{subfigure}%
    \begin{subfigure}[b]{.2\textwidth}
        \centering
        \begin{tikzpicture}[scale=0.62, font=\small,roundnode/.style={circle, draw=gray, fill=gray!5, very thick, minimum size=9mm, inner sep=0pt}]
    
            \node[roundnode, align=center] at (0,1.5) (a0) {a0};
            \node[roundnode, align=center] at (0,3) (a1) {a1};
            \node[roundnode, align=center] at (0,4.5) (a2) {a2};
            \node[roundnode, align=center] at (0,6) (a2a1) {a2,a1};
            
            \node[roundnode, align=center] at (2,1.5) (b0) {b0};
            \node[roundnode, align=center] at (2,3) (b2) {b2};
            \node[roundnode, align=center] at (2,4.5) (b3) {b3};
            \node[roundnode, align=center] at (2,6) (b3b2) {b3,b2};
    
            \draw[thick, -] (a0) -- (b0);
            \draw[thick, -] (a1) -- (b2);
            \draw[thick, -] (a1) -- (b3);
            \draw[thick, -] (a2) -- (b2);
            \draw[thick, -] (a2) -- (b3);
            \draw[thick, -] (a2a1) -- (b3b2);

        \end{tikzpicture}
        \caption{Potential Mergings}
        \label{fig:compatibility_graph_1}

    \end{subfigure}
    \hfill
    \begin{subfigure}[b]{.27\textwidth}
        \centering
        \begin{tikzpicture}[scale=0.55, font=\small,roundnode/.style={circle, draw=gray, fill=gray!5, very thick, minimum size=10mm,inner sep=0pt}]
            
            \node[] at (2,5.3) {$w=80$};
            \node[] at (6,5.3) {$w=80$};
            \node[] at (0,3.3) {$w=80$};
            \node[] at (6,-1.4) {$w=80$};
            \node[] at (4,3.4) {$w=12$};
            \node[] at (2,-1.5) {$w=30$};
            
            \node[text=blue, text width=1.5cm, align=center] at (-.5,0) {Maximum Weight Clique};
            
            \node[roundnode, align=center, draw=blue] at (0,2) (a1b2) {a1/b2};
            \node[roundnode, align=center, draw=blue] at (2,4) (a2b3) {a2/b3};
            \node[roundnode, align=center, draw=blue] at (2,0) (a2a1b3b2) {a2,a1/\\b3,b2};
            \node[roundnode, align=center, draw=blue] at (4,2) (a0b0) {a0/b0};
    
            \node[roundnode, align=center] at (6,4) (a1b3) {a1/b3};
            \node[roundnode, align=center] at (6,0) (a2b2) {a2/b2};
    
            \draw[ultra thick, -, blue] (a1b2) -- (a2b3);
            \draw[ultra thick, -, blue] (a0b0) -- (a2b3);
            \draw[ultra thick, -, blue] (a0b0) -- (a1b2);
            \draw[ultra thick, -, blue] (a0b0) -- (a2a1b3b2);
            \draw[ultra thick, -, blue] (a2b3) -- (a2a1b3b2);
            \draw[ultra thick, -, blue] (a1b2) -- (a2a1b3b2);
    
            \draw[thick, -] (a0b0) -- (a1b3);
            \draw[thick, -] (a0b0) -- (a2b2);
            \draw[thick, -] (a2b2) -- (a1b3);
            
        \end{tikzpicture}
        \caption{Compatibility Graph}
        \label{fig:compatibility_graph_2}

    \end{subfigure}%
    \begin{subfigure}[b]{.23\textwidth}
        \centering
        \begin{tikzpicture}[scale=0.5, font=\small,roundnode/.style={circle, draw=gray, fill=gray!5, very thick, minimum size=8.5mm,inner sep=0pt}, datashape/.style={trapezium, draw=gray, trapezium left angle=-65,trapezium right angle=-65, fill=gray!5, very thick}]
            \node[roundnode, align=center] at (0,4) (const) {const};
            \node[roundnode, align=center] at (1.5,2) (mul) {\large$\times$};
            \node[roundnode, align=center] at (4,2) (alu0) {\large$+$};
            \node[roundnode, align=center] at (2.3,-1.5) (alu1) {\large$+$};
    
            \node[datashape] at (1,0) (mux) {MUX};
            
            \draw[very thick, <-] (alu0.north west) -- ++ (-.3, +.6);
            \draw[very thick, <-] (alu0.north east) -- ++ (+.3, +.6);
            \draw[very thick, <-] (mul.north east) -- ++ (+.3, +.6);
            \draw[very thick, ->] (mul.south) -- (mux);
            \draw[very thick, ->] (alu0.south) -- (alu1.north east);
            \draw[very thick, ->] (const.south) -- (mul.north west);
            \draw[very thick, ->] (const.south) to [out=260,in=130] (mux);
            \draw[very thick, ->] (mux.south) -- (alu1.north west);
            
        \end{tikzpicture}
        \caption{Reconstructed Merged Graph}
        \label{fig:resulting_merged_graph}

    \end{subfigure}%
    \caption{Subgraph merging.}\label{fig:subgraph_merging}
    \vspace{-0.5cm}
    \end{centering}
\end{figure}
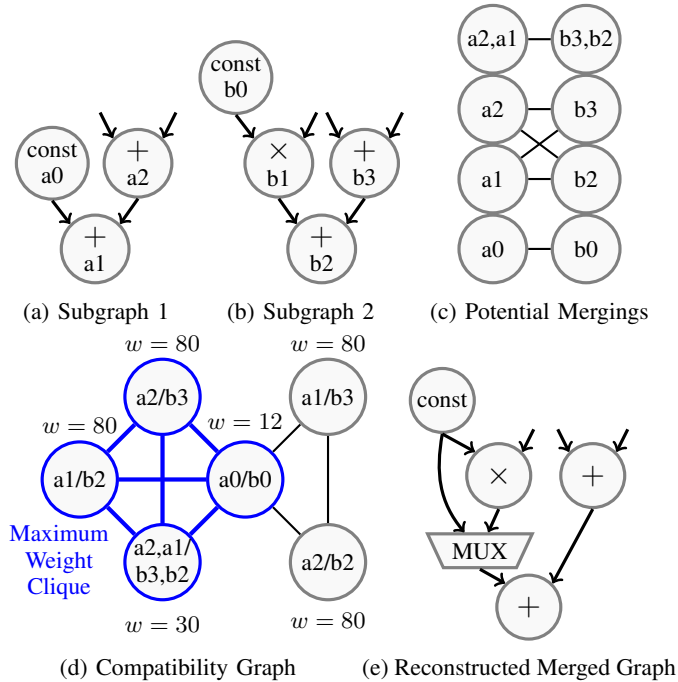

Merging many subgraphs from one or more applications allows for the acceleration of multiple distinct parts of one application or even multiple different applications using one PE architecture. However merging subgraphs is not a trivial problem. We have taken inspiration from a set of algorithms designed for high level synthesis for automated datapath graph merging \cite{merging}. The goal of these algorithms is to create a single structure that implements all of the distinct operations in the subgraphs with minimal area overhead. It produces one datapath that can be configured to each of the operations of each of the subgraphs. As an example, Fig.~\ref{fig:subgraph_1} and Fig.~\ref{fig:subgraph_2} show two subgraphs that we want to merge together.

\begin{figure*}
    \centering
    \includegraphics[width=1.0\textwidth]{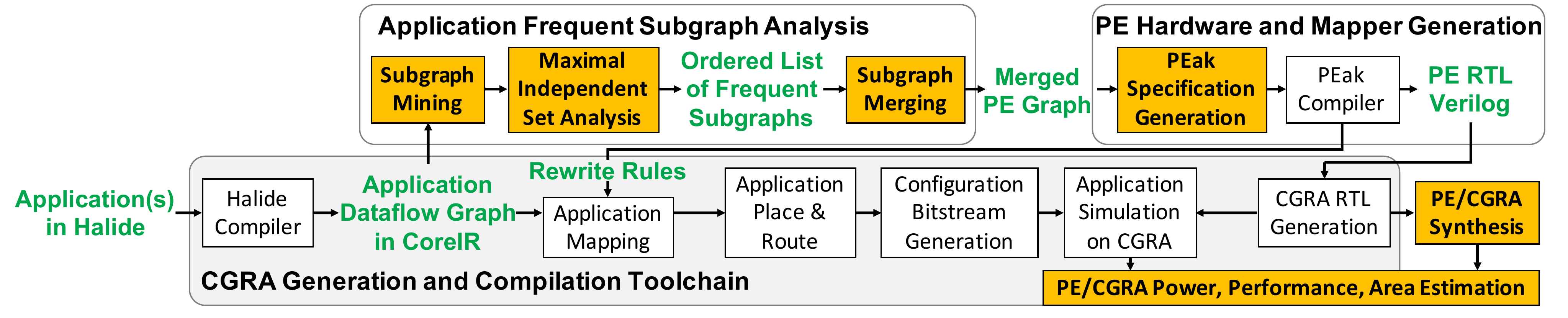}
    \caption{Our design space exploration framework. Steps from this work are in yellow. Intermediate representations are in green.}
    \label{fig:dse-flow}
    \vspace{-0.5cm}
\end{figure*}

The first step in the subgraph merging process is to create a set of potential merging opportunities between nodes of the same operation in each subgraph. Fig.~\ref{fig:compatibility_graph_1} shows a bipartite graph with nodes and edges in subgraph A on the left, and nodes and edges in subgraph B on the right. An edge from one node on the left to one on the right represents a potential merging opportunity. Two nodes can be merged if they are either the same operation, or can both be implemented on the same hardware block. Two edges can be merged if each of their endpoint nodes can be merged and the ports on the destination node match. Nodes and edges which do not have any potential merging opportunities in this example have been omitted from Fig.~\ref{fig:compatibility_graph_1} for simplicity. 

In this example, nodes $a0$ and $b0$ are both constants, so there is a corresponding edge in the bipartite graph. Nodes $a1$, $a2$, $b2$, and $b3$ are all add operations, so there are corresponding edges between these nodes. Finally, the edge $a2 \rightarrow a1$ and the edge $b3 \rightarrow b2$ start at an add operation and end at an add operation, and the ports on both $a1$ and $b2$ match, so those edges can be potentially merged as well. 

Next, these potential merging opportunities are transformed into a compatibility graph, where each potential merging is represented as a node, and each compatible merging is represented as an edge. Merging opportunities are compatible if they can both be implemented at the same time. Two mergings are incompatible if they merge one node in subgraph 1, to more than one nodes in subgraph 2, or vice versa. For example, merging $a1/b2$ is incompatible with merging $a2/b2$. 

Each node in the compatibility graph is given a weight, $w$, corresponding to the area reduction associated with applying the given merge. This area reduction is calculated by synthesizing the primitive nodes used in the subgraphs and calculating their area. For example, node $a1/b2$ represents merging $a1$ with $b2$. If this merging were applied, the resulting merged subgraph would only contain one adder for both of these nodes. The area reduction associated with this is the area of one adder.

To find the merging with the lowest area overhead, the maximum weight clique of this compatibility graph is calculated. The maximum weight clique of a graph is the set of fully connected nodes which have the largest sum of weights. In Fig.~\ref{fig:compatibility_graph_2} the maximum weight clique is the set of nodes highlighted in blue. Using the maximum weight clique of the compatibility graph, the lowest cost merging of the two subgraphs can be reconstructed. This resulting merged graph is shown in Fig.~\ref{fig:resulting_merged_graph}. Note that a multiplexer is added to enable multiple paths from nodes $a0$ to $a1$ and $b1$ to $b2$.

While this technique allows for efficiently merging frequent subgraphs from an application, there is still the question of which and how many subgraphs to merge. We can use the maximal independent set analysis to first identify the most interesting subgraphs mined from the application. The mined subgraphs are ranked by MIS size so that subgraphs that have many overlapping occurrences are considered last. Then we can use the number of subgraphs merged together as a tuning knob to adjust the specialization of the PE to the target application(s). This allows for automated design and generation of PEs, while still allowing the designer to control the generality and specialization of the CGRA.

\section{Design Space Exploration Framework}
\label{exploration-framework}

We build on top of an existing open-source agile hardware design flow~\cite{dac} to create specialized CGRAs with PEs generated from our application analysis. Our overall DSE framework is shown in Fig.~\ref{fig:dse-flow}. Fig.~\ref{fig:baseline-pe} shows the high-level architecture of the CGRAs we generate. The CGRA contains an array of PE and memory (MEM) tiles connected through a statically configured interconnect containing horizontal and vertical routing tracks.

Our DSE framework (Fig.~\ref{fig:dse-flow}) has the following steps: 
\begin{enumerate}
    \item We start with applications from a domain written in Halide \cite{halide2013_journal}, a DSL for succinctly describing image processing and machine learning applications. 
    \item The Halide compiler from~\cite{dac} converts the Halide application into a dataflow graph in CoreIR~\cite{coreir}. 
    \item The application analysis flow (Section~\ref{application-analysis}) performs subgraph mining and maximal independent set analysis and generates an ordered list of frequent subgraphs. 
    \item Subgraph merging merges several frequent subgraphs to generate a candidate PE graph, which we automatically convert into a PE specification in PEak~\cite{dac} DSL. 
    \item The PEak compiler generates the PE Verilog and the rewrite rules required by the application mapper. The PE RTL is fed into CGRA RTL generation to generate the final CGRA hardware.
    \item Meanwhile, the application mapper, using the rewrite rules, generates a covering of the application CoreIR graph using PEs, while trying to minimize the number of PEs used. Depending on its architecture, a PE can execute a single operation or a small graph of operations. The result is a graph of PEs and MEMs.
    \item From this mapped graph, we generate the CGRA configuration bitstream and simulate the CGRA RTL using Synopsys VCS.
    \item Finally, we synthesize the CGRA Verilog in TSMC 16 nm technology and evaluate area and power for the PE and the CGRA using Synopsys Design Compiler and Synopsys PrimeTime PX.
\end{enumerate}

\section{Results}
\label{results}

\begin{figure}
    \centering
    \includegraphics[width=0.47\textwidth]{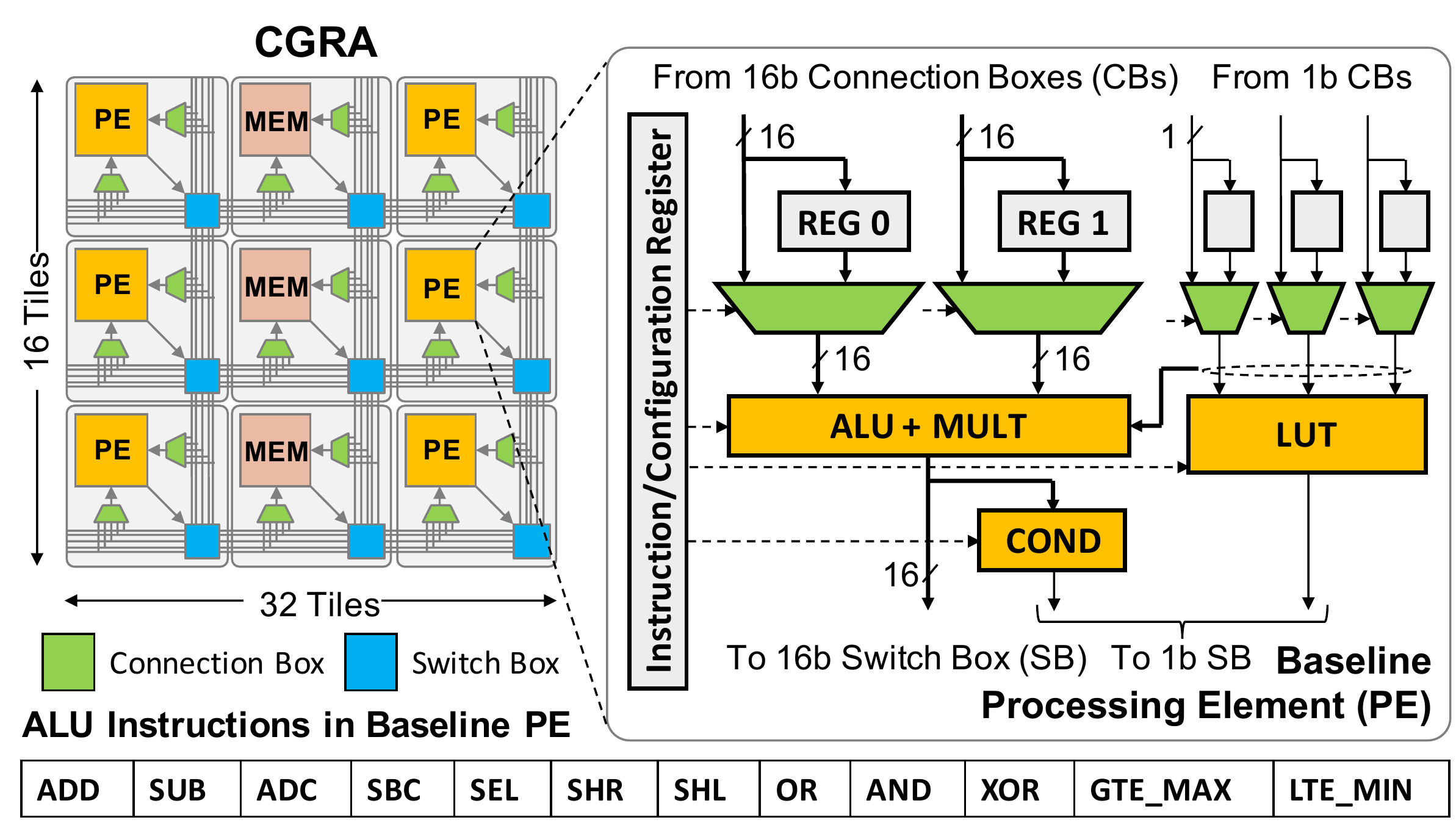}
    \caption{Baseline PE and CGRA architecture.}
    \label{fig:baseline-pe}
    \vspace{-0.5cm}
\end{figure}

This section presents the improvements we achieve by exploring the PE design space and specializing the PE architecture for a specific application or an application domain. Fig.~\ref{fig:baseline-pe} shows the architecture of the baseline PE from~\cite{dac} that we compare with. It contains an integer arithmetic unit and can perform bit operations using a look-up table (LUT). It is a very general PE that can execute most applications and is not specialized for any particular domain. Using our framework, we generate the following PE variations for each application by indicating which subgraphs are merged together: 
\begin{itemize}
    \item PE 1: The first PE variation is the baseline PE but with only the operations necessary for the application.
    \item PE 2: The second variation merges the subgraph with the largest maximal independent set with PE 1.
    \item PEs 3 - 5: Further variations additionally merge other subgraphs into the PE architecture in the order of their MIS size. The last variation for an application is the most specialized PE possible without increasing area or energy.
\end{itemize}  

\subsection{DSE for Image Processing Applications}

\begin{figure}
    \centering
        \includegraphics[width=0.245\textwidth]{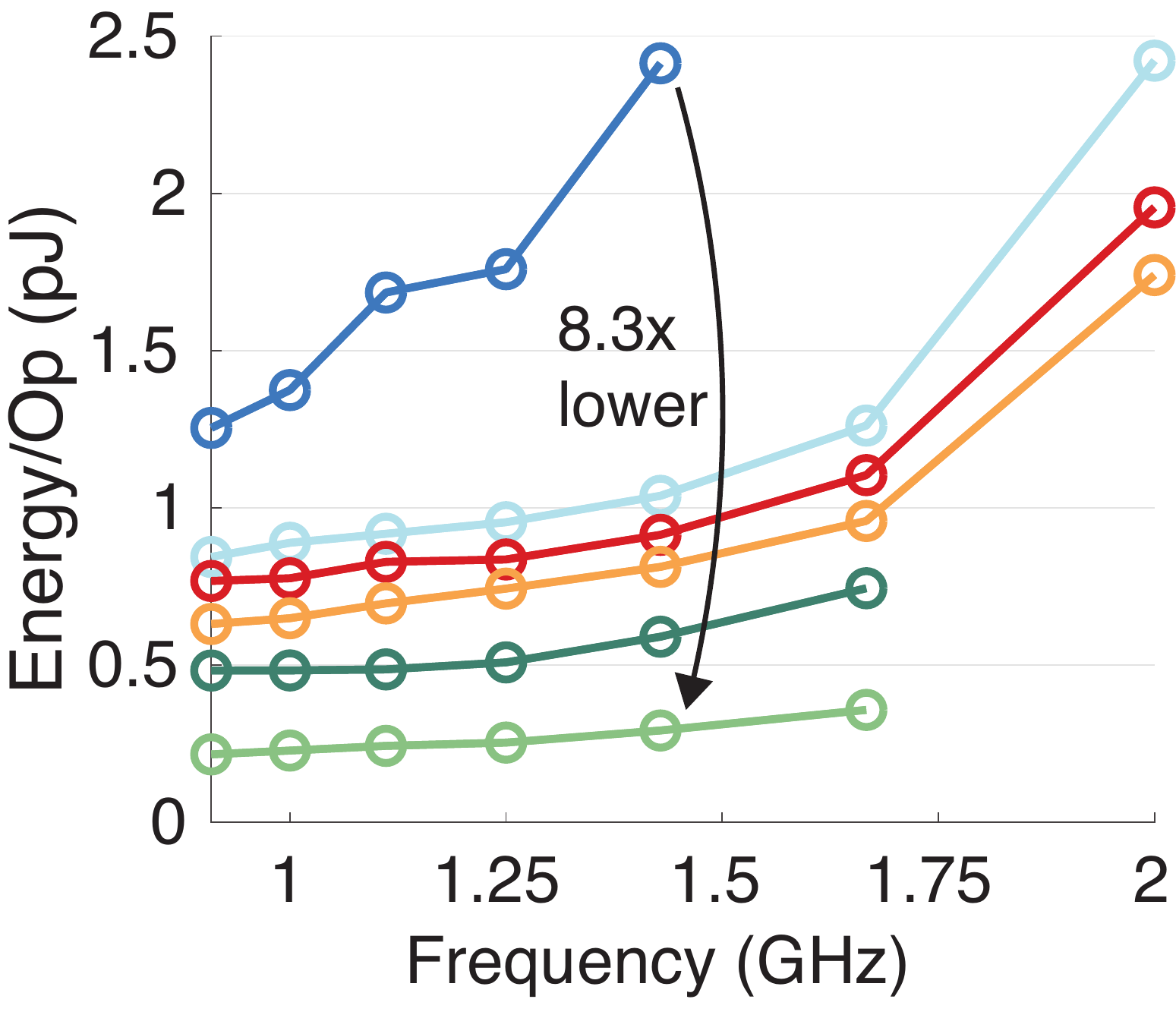}%
        \includegraphics[width=0.245\textwidth]{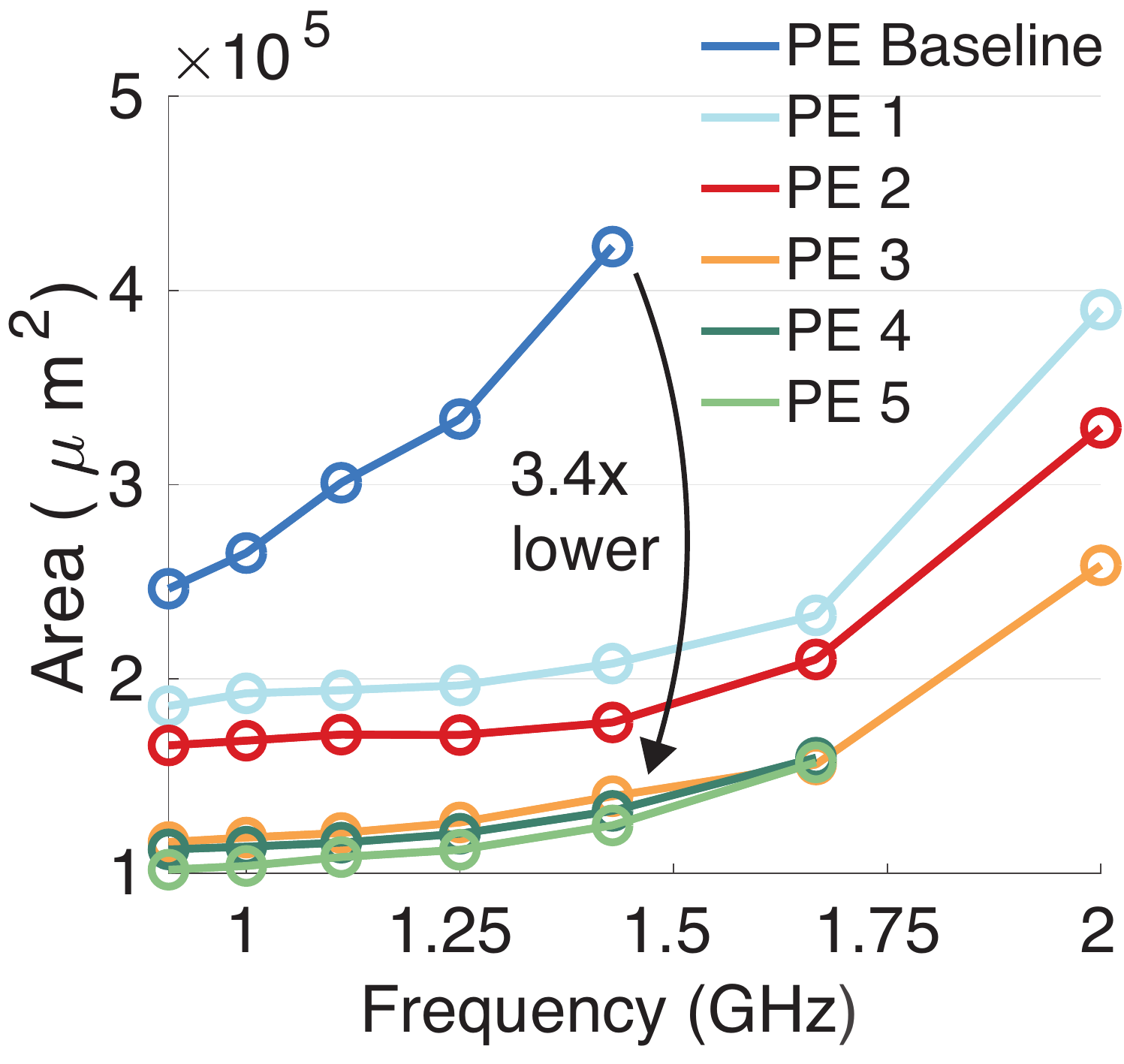}
    \caption{Energy per operation (op) and total area consumed by active PE cores as the CGRA is increasingly specialized for camera pipeline.}
    \label{fig:camera-pipeline-sweep}
    \vspace{-12pt}
\end{figure}

\begin{figure}
    \centering
    \includegraphics[width=0.5\textwidth]{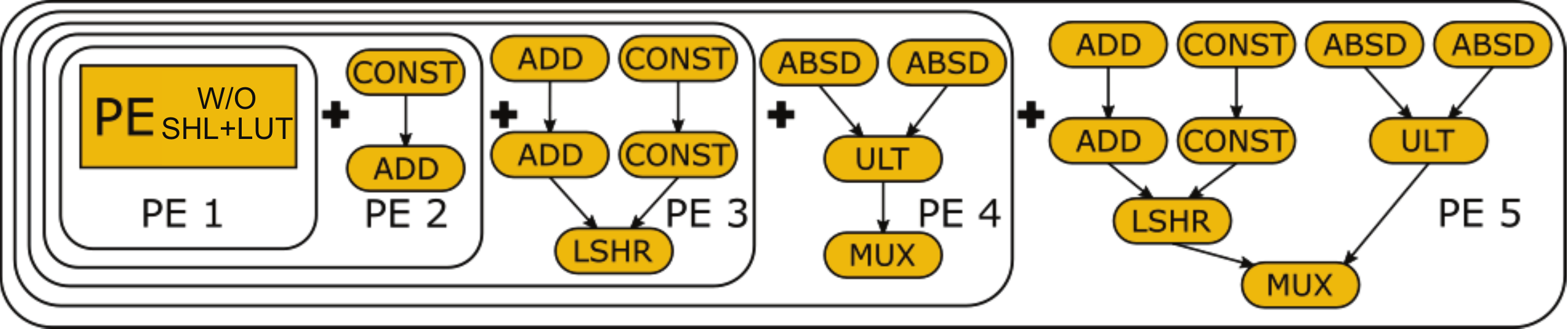}
    \caption{Subgraphs merged together to form PE variants 1 through 5 for camera pipeline.}
    \label{fig:camera-pipeline-subgraphs}
    \vspace{-10pt}
\end{figure}

\begin{figure}
    \centering
    \includegraphics[width=0.248\textwidth]{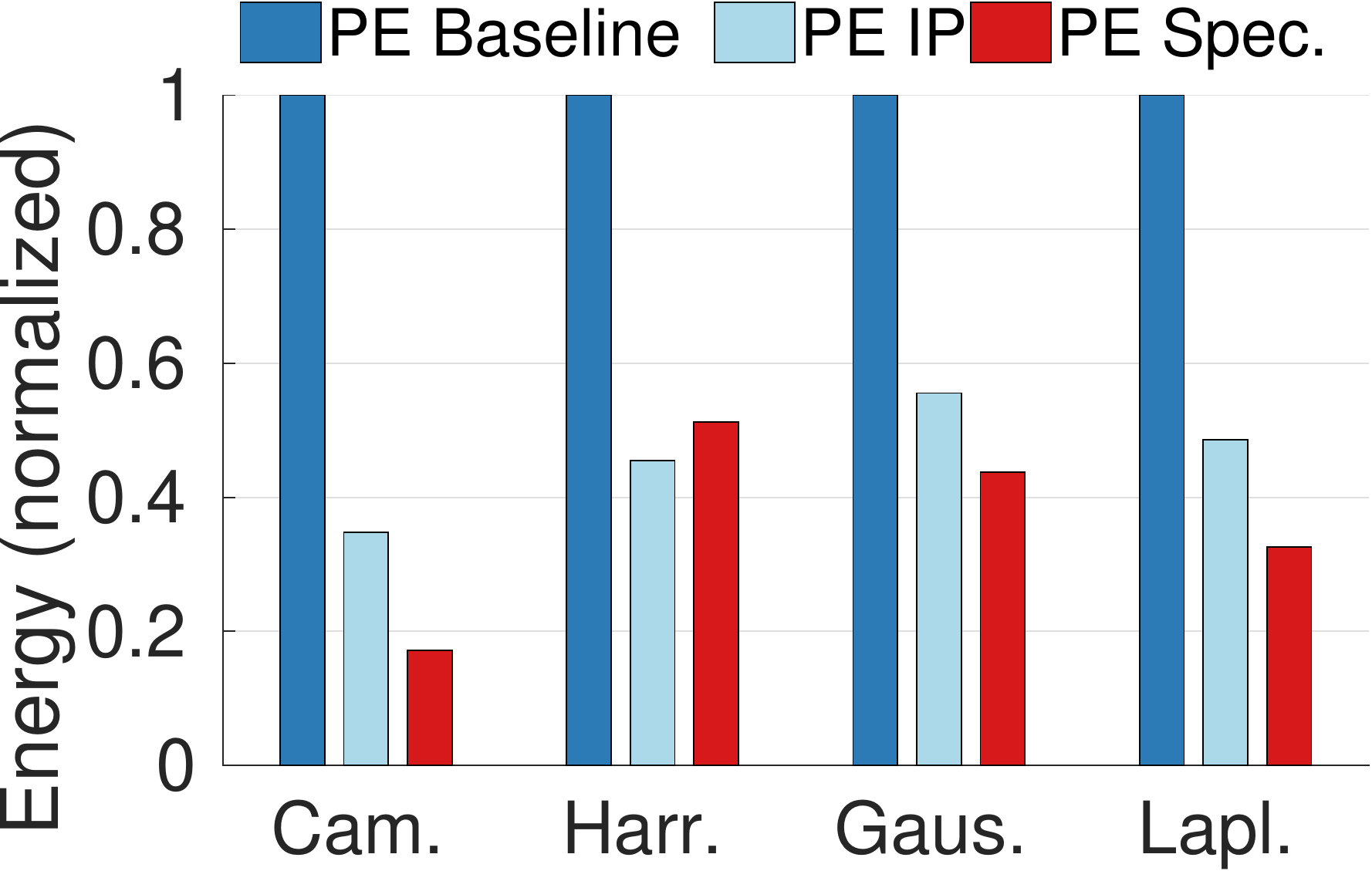}%
    \includegraphics[width=0.248\textwidth]{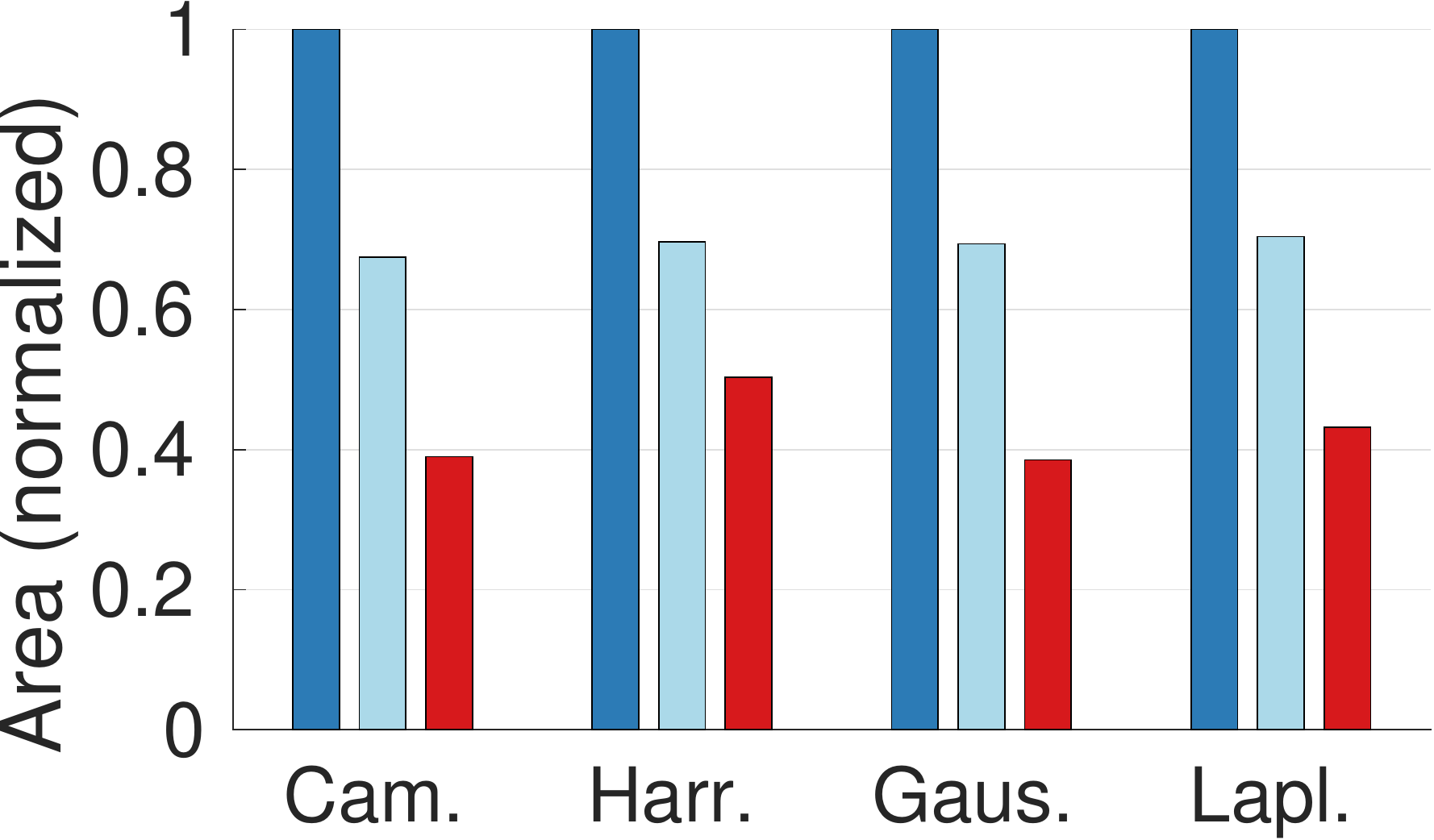}
    \caption{Normalized PE core energy and total area for all image processing applications run on a PE specialized for all four applications (PE IP) and a PE specialized for the particular application (PE Spec.).}
    \label{fig:all-image-processing}
    \vspace{-15pt}
\end{figure}

We focus on specializing across four imaging processing applications: Harris corner detection, Gaussian blur, camera pipeline and Laplacian pyramid.

We first analyze camera pipeline, which is the most complex application of the four. It uses all the operations in the baseline PE except for left shift (SHL) and bitwise logical operations (LUT) and needs 221 operations to compute an output pixel. Fig.~\ref{fig:camera-pipeline-subgraphs} shows the most frequent subgraphs and the architectures of the different PE variations for camera pipeline. Fig.~\ref{fig:camera-pipeline-sweep} shows the energy per operation dissipated by the PE core and total area (PE core area $\times$ number of PEs used by the application) of the different PE variants specialized for camera pipeline, swept across different synthesis frequencies.  Specializing the PE for camera pipeline results in up to $8.3\times$ and $3.4\times$ less energy and area, respectively, on the CGRA than the baseline PE. Performance also benefits from specialization; the baseline PE has a maximum frequency of 1.43 GHz while the camera pipeline-specific PEs can operate up to 2 GHz.

Further, by merging in frequent subgraphs from all four applications we create PE IP, a PE specialized for image processing. In Fig.~\ref{fig:all-image-processing}, PE IP supports all four applications, while PE Specialized (PE Spec.) is the most efficient PE created for the particular application (i.e. one of PEs 3-5). Optimizing the PE for all image processing applications (PE IP) results in a 29.6\% to 32.5\% decrease in PE area and 44.5\% to 65.25\% lower energy across all four applications. PE Spec. typically yields more benefits than PE IP; targeting the CGRA only for Laplacian pyramid, for example, further decreases area and energy from PE IP by 38.6\% and 33.0\%, respectively. It is interesting to note that PE IP is more energy-efficient than PE Spec. for Harris; in this instance PE IP has an architecture that reduces activity on an input to a multiplier. In general though, PEs from cross-application analysis are worse than application-specific PEs but significantly better than the baseline.

\subsection{DSE for Machine Learning Applications}

\begin{figure}
    \centering
    \includegraphics[width=0.248\textwidth]{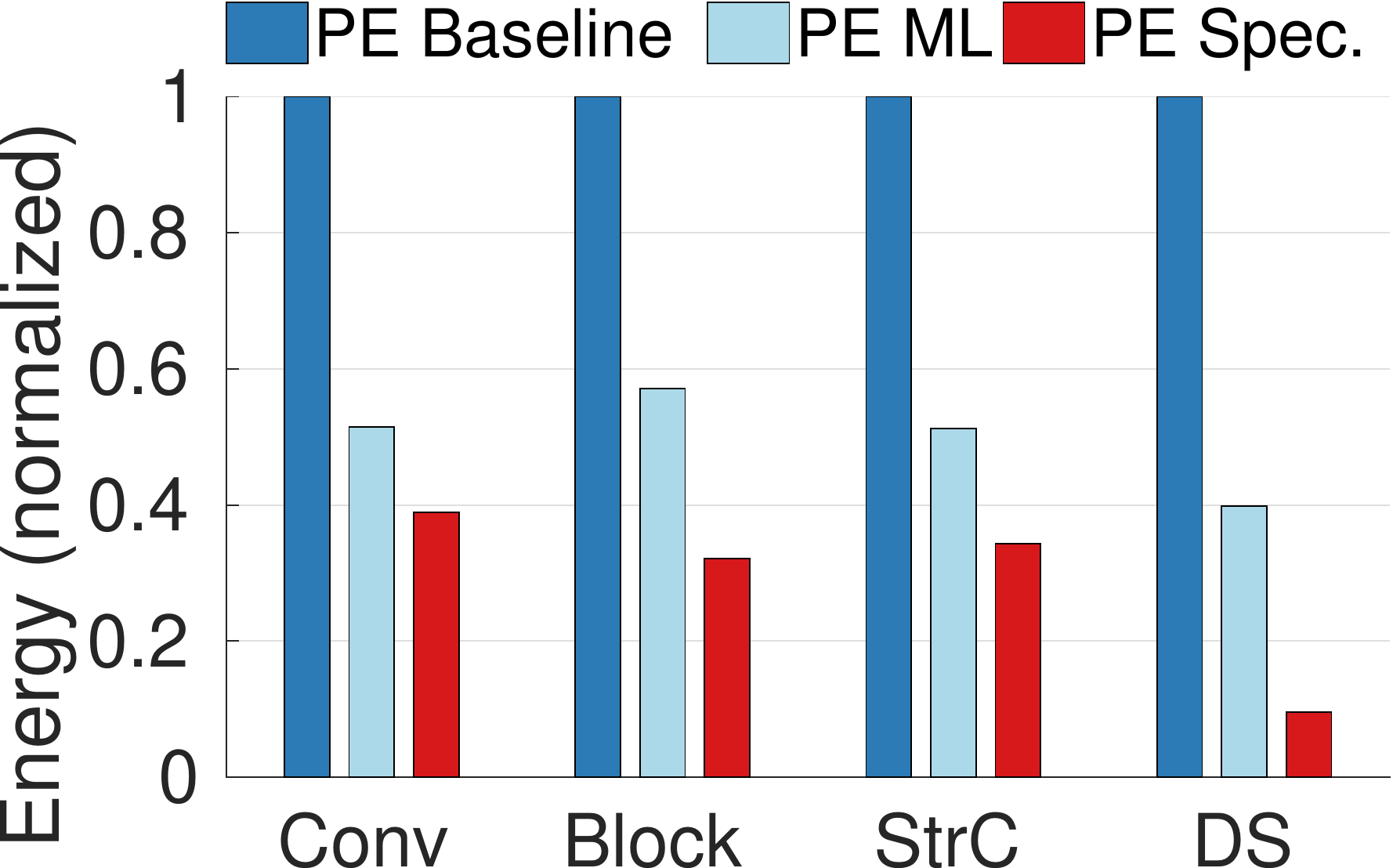}%
    \includegraphics[width=0.248\textwidth]{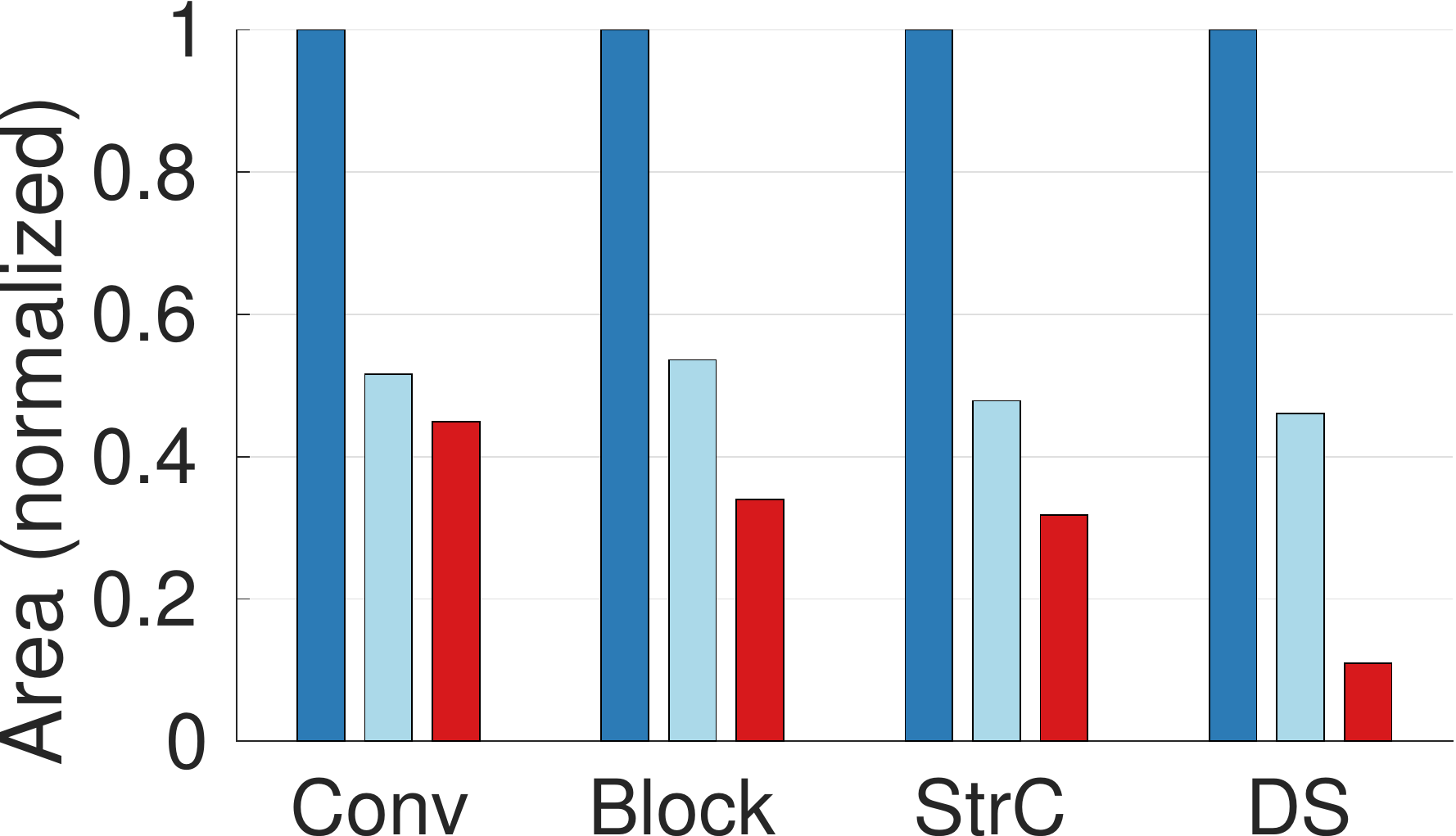}
    \caption{Normalized energy and area for ML kernels run on a PE specialized for ML (PE ML) and a PE specialized for the particular kernel (PE Spec).}
    \label{fig:all-ml}
\end{figure}

To develop a generic ML accelerator from the baseline CGRA, we analyze two popular networks, ResNet-50 and U-Net. Fig.~\ref{fig:all-ml} shows the effect of specializing the PE to common kernels found in both networks, including multichannel convolution (Conv), residual block (Block), strided convolution (StrC), and down sample (DS), and creating a generalized PE for ML (PE ML). PE ML, shown in Fig.~\ref{fig:pe-ml}, is a PE that efficiently computes convolutions but also handles shifts, general multiplication and add, and ReLUs. While PE ML is worse than the PE specialized for a lone kernel, it is able to run more applications while still consuming up to 60.15\% less energy than the baseline PE.

\begin{table}
  \begin{minipage}{0.5\textwidth}
  \begin{minipage}[b]{0.25\textwidth}
    \centering
    \includegraphics[width=0.97\textwidth]{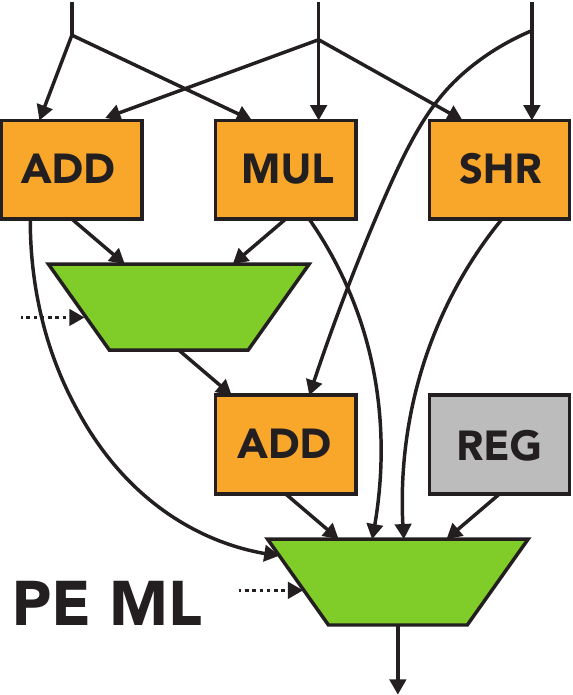}
    \captionof{figure}{}
    \label{fig:pe-ml}
  \end{minipage}%
  \begin{minipage}[b]{0.7\textwidth}
    \centering
    \footnotesize
    \begin{tabular}{c c c c} \toprule
        & Simba \cite{simba} & CGRA & ML CGRA \\ \midrule
        PE & Vector MACs & Baseline & PE ML \\
        Tech. & 16nm & 16nm & 16nm \\
        Precision & 8b & 16b & 16b \\
        MHz & 161-2001 & 909 & 909 \\
        pJ/op & 0.105-2.15 & 2.25 & 1.75 \\
        \bottomrule
    \end{tabular}
    \captionof{table}{Comparison of ML CGRA with an ASIC (Simba) and a generic CGRA.}
    \label{tab:ml-comparison}
    \end{minipage}
  \end{minipage}
  \vspace{-12pt}
\end{table}

Table~\ref{tab:ml-comparison} shows our CGRAs compared against Simba \cite{simba}, a state-of-the-art ML accelerator. The PEs in Simba have eight 8-bit precision vector MAC units, and the energy efficiency of Simba is derived from its reported peak and ResNet-50 performance. After factoring in the cost of the memory tiles on the CGRA, specializing the PEs reduces overall energy by 22.1\% over a baseline CGRA and results in a CGRA design that nears the efficiency of a custom accelerator.

\section{Related Work}

Automated accelerator generation for specific application domains has been developed previously, particularly for neural networks. FPGA-based automated generators for neural networks include DnnWeaver \cite{dnnweaver} and DNNBuilder \cite{dnnbuilder}. MAGNet~\cite{magnet} is an accelerator generator for neural networks that explores different application mappings for hardware-software co-optimization. However, it generates ASICs and not CGRAs, does not consider applications outside of neural networks, and does not perform cross-application analysis. 

Exploring the architecture of CGRAs has also been previously studied. Expression-Grained Reconfigurable Arrays (ERGAs)~\cite{egra} explore hardware tradeoffs in CGRA computing elements using clusters of ALUs. However, the DSE is not guided by application analysis. Plasticine~\cite{plasticine} is a parameterizable CGRA with interleaved compute units and memory units in an interconnect, though automated design space exploration is not included in this work.

General automated DSE frameworks have also been studied in the past. HyperMapper~\cite{kunle-dse} is a DSE framework intended to do optimization for complex design spaces. While this technique does an automated search, it is guided by a hardware description and not by analysis of applications.

\section{Conclusion}
\label{conclusion}
We have presented a design space exploration framework that allows for automated specialization of CGRA PEs to an application or an application domain, enabling the creation of complex, high-performance PEs that lower overall energy and area costs. We demonstrate that specializing PEs results in up to $9.1\times$ and $10.5\times$ reduction in area and energy, respectively, across a variety of image processing and machine learning applications, and a CGRA with specialized PEs approaches the efficiency of a domain-specific accelerator.

\bibliographystyle{plain}
\bibliography{references}

\end{document}